\documentclass[useAMS,usenatbib]{mn2e}

\usepackage{graphicx}
\usepackage{verbatim}   % for comments
\usepackage{times}

\newcommand{\vtbold}[1]{#1}               % for arXiv
\title[Period-luminosity relations of pulsating M giants]{Period-luminosity relations of pulsating M giants in the solar neighbourhood and the Magellanic Clouds}
\author[V. Tabur et al.]{V. Tabur$^{1}$\thanks{E-mail: tabur@physics.usyd.edu.au}, T.R. Bedding$^1$, L.L. Kiss$^{1,2}$, T. Giles$^1$, A. Derekas$^{1,2,3}$ and T.T. Moon$^1$\\
$^{1}$Sydney Institute for Astronomy (SIfA), School of Physics, The University of Sydney, NSW 2006, Australia\\
$^{2}$Konkoly Observatory, Hungarian Academy of Sciences, PO Box 67, H-1525 Budapest, Hungary\\
$^{3}$Magyary Fellow, Dept. of Astronomy, E\"otv\"os University, Budapest, Hungary\\
}

\begin{document}

\date{Accepted 2010 July 9. Received 2010 July 7; in original form 2010 February 6}
\pagerange{\pageref{firstpage}--\pageref{lastpage}} \pubyear{2010}
\maketitle
\label{firstpage}

\begin{abstract}
We analyse the results of a 5.5-yr photometric campaign that monitored 247 southern, semi-regular variables with relatively precise Hipparcos parallaxes %. These data extend the period-luminosity (P-L) relationships in the ($K$, $\log P$) plane to lower luminosities and shorter periods than previously observed, 
\vtbold{to} demonstrate an unambiguous detection of Red Giant Branch (RGB) pulsations in the solar neighbourhood. We show that Sequence A$'$ contains a mixture of AGB and RGB stars, as indicated by a temperature related shift at the TRGB. Large Magellanic Cloud (LMC) and Galactic sequences are compared in several ways to show that the P-L sequence zero-points have a negligible metallicity dependence. We describe a new method to determine absolute magnitudes from pulsation periods and calibrate the LMC distance modulus using Hipparcos parallaxes to find $\mu_{\rm LMC} = 18.54 \pm 0.03$ mag. Several sources of systematic error are discussed to explain discrepancies between the MACHO and OGLE sequences in the LMC. We derive a relative distance modulus of the Small Magellanic Cloud (SMC) relative to the LMC of $\Delta \mu = 0.41 \pm 0.02$ mag. A comparison of other pulsation properties, including period-amplitude and luminosity-amplitude relations, confirms that RGB pulsation properties are consistent and universal, indicating that the RGB sequences are suitable as high-precision distance indicators. \vtbold{The M giants with the shortest periods bridge the gap} %, both in luminosity and period,
between G and K giant solar-like oscillations and M-giant pulsation, revealing a smooth continuity as we ascend the giant branch.
\end{abstract}


%%%%%%%%%%%%%%%%%%%%%%%%%%%%%%%%%%%%%%%%
\section{Introduction}
%%%%%%%%%%%%%%%%%%%%%%%%%%%%%%%%%%%%%%%%

The existence of multiple period-luminosity (P-L) relationships is now well-established among Miras and semi-regular variables (SRVs) in the Large Magellanic Cloud (LMC), Small Magellanic Cloud (SMC), and Galactic Bulge. In the last decade, microlensing surveys have provided a wealth of information, resulting in a completely new picture of red giant pulsations \citep[see, for example,][]{b_woo99,b_kis03,b_kis04,b_ita04,b_gro04,b_sch,b_wra,b_sos07}.
Most studies have concentrated on the ($K$, $\log P$) plane, where M-giant emissions are strongest and interstellar extinction is relatively low, although similar P-L sequences have recently been shown for the mid-IR \citep{b_gla09}.

M giants are common and intrinsically luminous, making them potentially important as distance indicators. Indeed, their P-L relations have already been used to measure the distances to the external galaxies NGC 5128 \citep{b_rej04} and the Phoenix Dwarf \citep{b_men}, to determine the relative distance modulus between the LMC, SMC, and Bulge \citep{b_kis04,b_gla09}, and to probe the three-dimensional structure of the LMC \citep{b_lah05}.
However, the metallicity dependence of the P-L relations is still uncertain, limiting their usefulness. To determine whether the P-L relations are truly universal, we must first identify the local sequences in the solar neighbourhood.

The first evidence of multiple local sequences was provided by \citet{b_bed98}, who used Hipparcos parallaxes and published periods for 24 nearby M giants to demonstrate two distinct sequences.
Subsequent investigations by \citet{b_kna} and \citet{b_yes} confirmed at least two sequences, but these studies were limited by relatively large parallax errors, and a reduced sample size (by tightly constraining parallax errors), respectively. Recently, \citet{b_gla07} used revised Hipparcos parallaxes and published periods for 64 red giants, confirming that local stars obey similar P-L relations to those found in the Magellanic Clouds and Bulge, although details remained sketchy due to insufficient data, particularly for short-period stars.

In Paper I \citep{b_tab09a}, we provided periods and $K$-band magnitudes for 247 nearby \vtbold{red} giants that we observed over 5.5 years. With the benefit of revised Hipparcos parallaxes for this large sample, we now analyse those data to study the local sequences.
We describe our sample in Sect. \ref{s_sample}, and provide period-luminosity diagrams in Sect. \ref{s_PL}. The sequences are compared to those in the LMC to determine the effect of metallicity on the P-L zero-point (Sect. \ref{s_lmc} \& \ref{s_pl_zp}). A new method to estimate absolute magnitudes from periods alone is described in Sect. \ref{s_ruler}, and is used to calibrate the LMC distance modulus using Hipparcos parallaxes. We compare pulsation amplitudes to those in previous studies (Sect. \ref{s_Amp}) and demonstrate how our sample relates to G and K giants that exhibit solar-like oscillations (Sect. \ref{s_SLO}), and summarise our conclusions in Sect. \ref{s_conclusion}.

%%%%%%%%%%%%%%%%%%%%%%%%%%%%%%%%%%%%
\section{Data}
\label{s_sample}
%%%%%%%%%%%%%%%%%%%%%%%%%%%%%%%%%%%

% 247 w/periods - 4 (3C+1S) = 243
Our sample consists of 243 nearby M giants\footnote{We have excluded 4 stars with spectral types C or S from the original list of 247 stars with periods published in Paper I.} with periods and absolute $K$ magnitudes determined in Paper I, supplemented by 64 additional stars tabulated in Table 2 of \citet{b_gla07}. We used our periods in preference to the literature values for 21 stars that were in common.

The periods derived from CCD observations in Paper I were limited to $P < 300$ d, since we considered those to be the most reliable. Longer periods were less well-determined due to gaps in the sampling, possible instrumental drifts, and difficulty matching light curves in different survey phases. However, periods derived using photoelectric photometry and those extracted from \citet{b_gla07} were retained for all values of the period.

Apparent $K$-band magnitudes were either determined from the best available NIR catalogue, as described in Paper I, or obtained directly from Table 2 of \citet{b_gla07}. For HD 201298, the value of $m_{K_s} = 1.06$ mag listed by \citet{b_gla07} is inconsistent with the value given in the Two-Micron All-Sky Survey (2MASS; \citealt{b_cut}), namely $m_{K_s} = 1.729 \pm 0.248$ mag, and we have adopted the latter value. The use of revised Hipparcos parallaxes \citep{b_van07} resulted in a reduction of the median relative parallax uncertainty of the sample from $\sigma_{\pi}/\pi$ = 0.175 to 0.106, resulting in a median uncertainty in $M_K$ of 0.23 mag. To sharpen the P-L ridges, we restricted our analysis to a subset of stars with the smallest relative parallax uncertainties, as described in Sect. \ref{s_PL}. However, before discussing the sequences of periods, we first present examples of the power spectra from which they were extracted.

%%%%%%%%%%%%%%%%%%%%%%%%%%%%%%%
\subsection{Power spectra}
%%%%%%%%%%%%%%%%%%%%%%%%%%%%%%%

\begin{figure}
 \includegraphics[scale=1.0, angle=0]{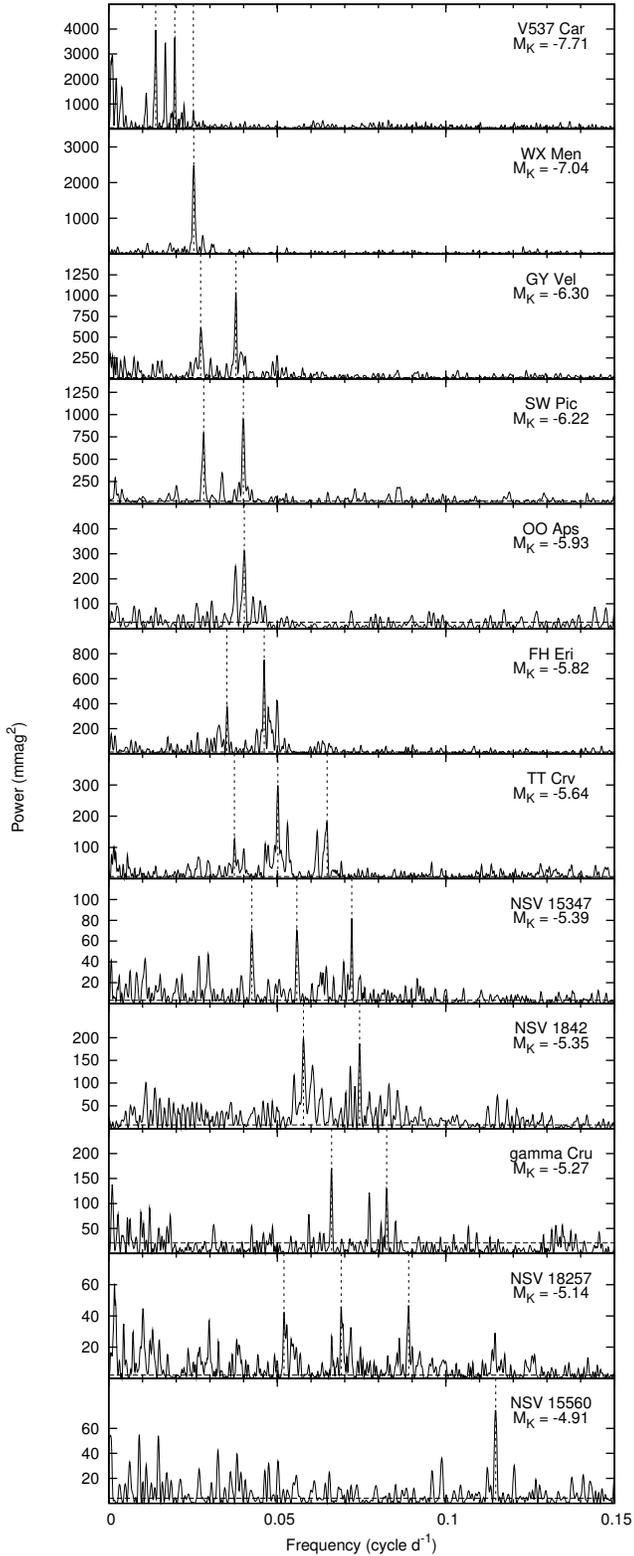}
 \caption{A selection of power spectra ordered by decreasing absolute luminosity that demonstrate the trend toward low-amplitude, short-period pulsation with decreasing luminosity, giving the visual impression of sequences running diagonally down the plot. Significant periods (selected from Paper I) are marked with dotted vertical lines. Dashed horizontal lines indicate the mean signal in the frequency range 0.2--0.4 cycle d$^{-1}$.}
 \label{fig015}
\end{figure}

Figure~\ref{fig015} shows a selection of Fourier spectra ordered by decreasing $M_K$. The ordinate shows the power, being the square of the Fourier semi-amplitude. The overall trend toward shorter periods and smaller amplitudes with decreasing luminosity is clear (note the change in vertical scale). Also visible is the correlation between adjacent spectra, where the dominant peaks are slowly shifting toward shorter periods as luminosity decreases and, in some cases, less prominent, shorter-period peaks start to dominate (see, for example, the transition from FH Eri to TT Crv, and NSV 18257 to NSV 15560). Noise increases toward the bottom of the plot, where lower amplitudes mean the photometric errors become more significant.

% Fig 3.28, 072/x.png

% Fig 3.xx / 087
\begin{figure}
 \includegraphics[scale=1.0, angle=0]{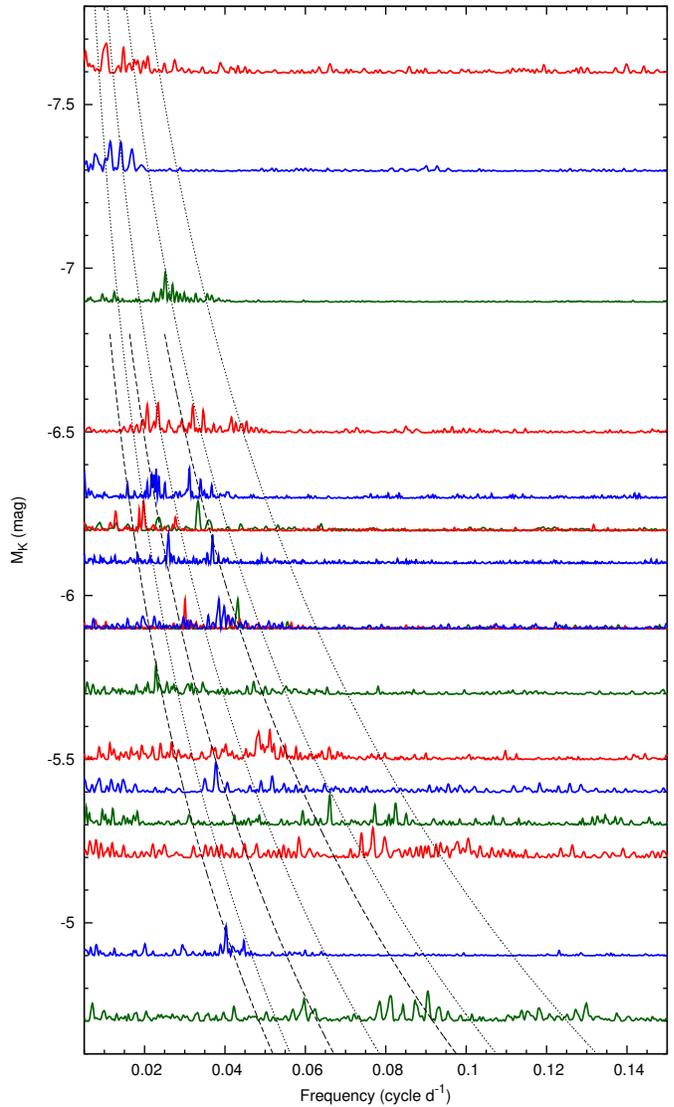}
 \caption{Normalised power spectra of stars with $\sigma_{\pi}/\pi \le 0.04$ and $M_K$ rounded to 0.1 mag. LMC sequences from S07 are shown for comparison, assuming $\mu_{\rm LMC}=18.50$. Dashed (dotted) lines represent Sequences $b1$--$b3$ ($a1$--$a4$) respectively.}
 \label{fig017}
\end{figure}

Figure~\ref{fig017} shows normalised power spectra for the subset of stars with the most precise parallaxes ($\sigma_{\pi}/\pi \le 0.04$), with $M_K$ rounded to the nearest 0.1 mag to avoid overlapping spectra. Dotted lines indicate the positions of the P-L sequences using the nomenclature of \citet[hereafter S07]{b_sos07}. These plots demonstrate a clear detection of the P-L sequences, and the prevalence of short-period pulsation at low luminosities.

%%%%%%%%%%%%%%%%%%%%%%%%%%%%%%%%%%%%%%%%%%%%%%
\section{Period-Luminosity Relations}
\label{s_PL}
%%%%%%%%%%%%%%%%%%%%%%%%%%%%%%%%%%%%%%%%%%%%%%

% Fig 3.14 / 076
\begin{figure}
 \includegraphics[scale=1.0, angle=0]{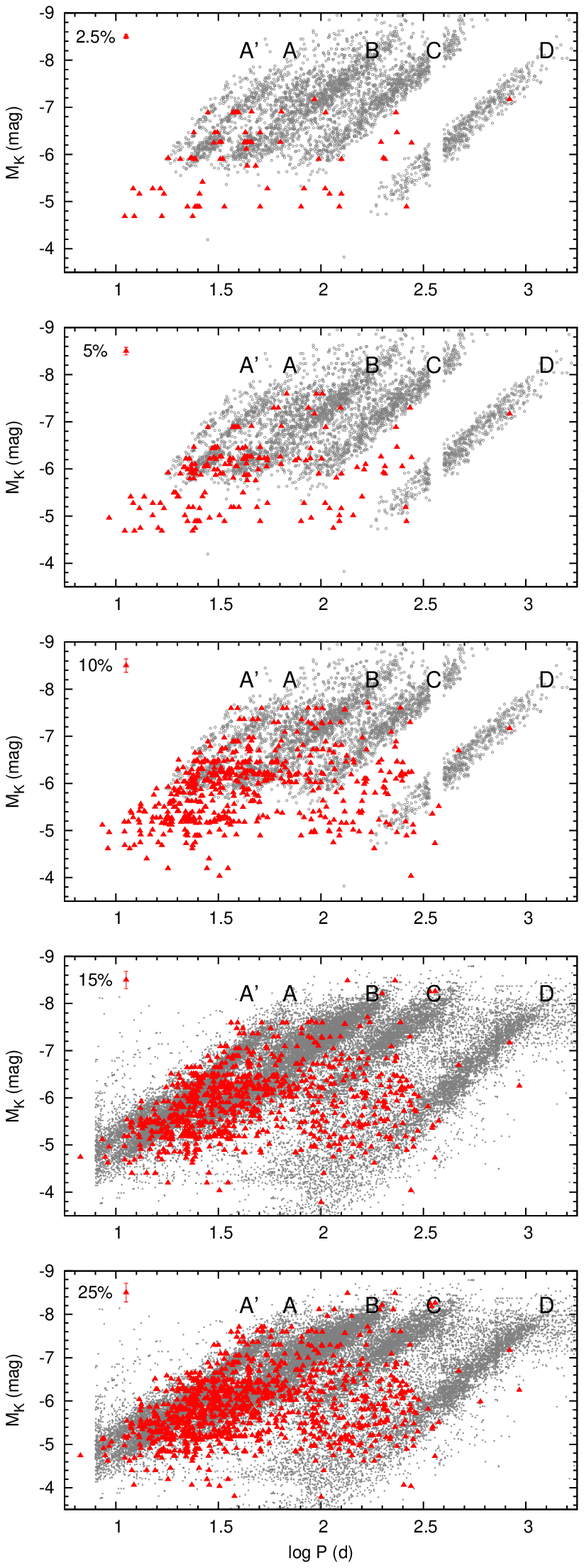}
 \caption{Period-Luminosity diagram of Galactic SRVs with $\sigma_\pi / \pi \le 0.025, 0.05, 0.10, 0.15, 0.25$ (red triangles) and LMC stars (grey dots), with an assumed LMC distance modulus of 18.50. \vtbold{The top 3 panels use MACHO periods from D06. The bottom 2 panels use OGLE periods from S07, and show the overlapping luminosity range.}}
 \label{fig001}
\end{figure}

Figure~\ref{fig001} shows period-luminosity diagrams in the ($M_K$, $\log P$) plane for Galactic SRVs with five different thresholds on the parallax uncertainties (triangles). For comparison, we have plotted the corresponding LMC sequences from Massive Compact Halo Object (MACHO) survey data analysed by \citet[][hereafter D06]{b_der}, assuming a distance modulus of $\mu_{\rm LMC} = 18.50$ \citep{b_alv04}. \vtbold{We show \vtbold Optical Gravitational Lensing Experiment (OGLE) data from S07 for relative parallax errors larger than 10 per cent.} The sequences are labelled A$'$, A, B, C and D, following the nomenclature established by \citet{b_woo99}. Sequence A$'$, first identified by \citet{b_sos04}, consists of the shortest periods and smallest amplitudes, and is well-resolved in the \vtbold{sparser} MACHO data.

We briefly note a number of features in Figure~\ref{fig001} and discuss them in greater detail in following sections. Firstly, there is broad agreement between our Galactic sample and the LMC sequences, giving an overall impression of parallel ridges in both datasets. The agreement is excellent at small $\sigma_{\pi} / \pi$, particularly for Sequences A$'$ and A, which are relatively narrow. The Galactic sequences becomes less clear as parallax restrictions are relaxed, presumably due to relative parallax uncertainties, which contribute up to $\sim$ 0.5 mag of scatter. Error bars have been omitted for clarity, but we show the median uncertainty in each panel.

Stars with periods longer than 300 d ($\log P \sim 2.5$) were obtained from \citet{b_gla07}. A few have periods longer than those on Sequence D, and may be spurious. Periods that fall between Sequences C and D we attribute to low-frequency stellar noise, presumably from convection \citep{b_kis06}.

The significant decrease in stellar density at luminosities brighter than $M_K \sim -6.8$ marks the tip of the Red Giant Branch (TRGB), consistent with the recent finding that the local TRGB is detectable using revised Hipparcos parallaxes and NIR photometry from 2MASS and DIRBE \citep{b_tab09b}.

\begin{figure}
 \includegraphics[scale=1.0, angle=0]{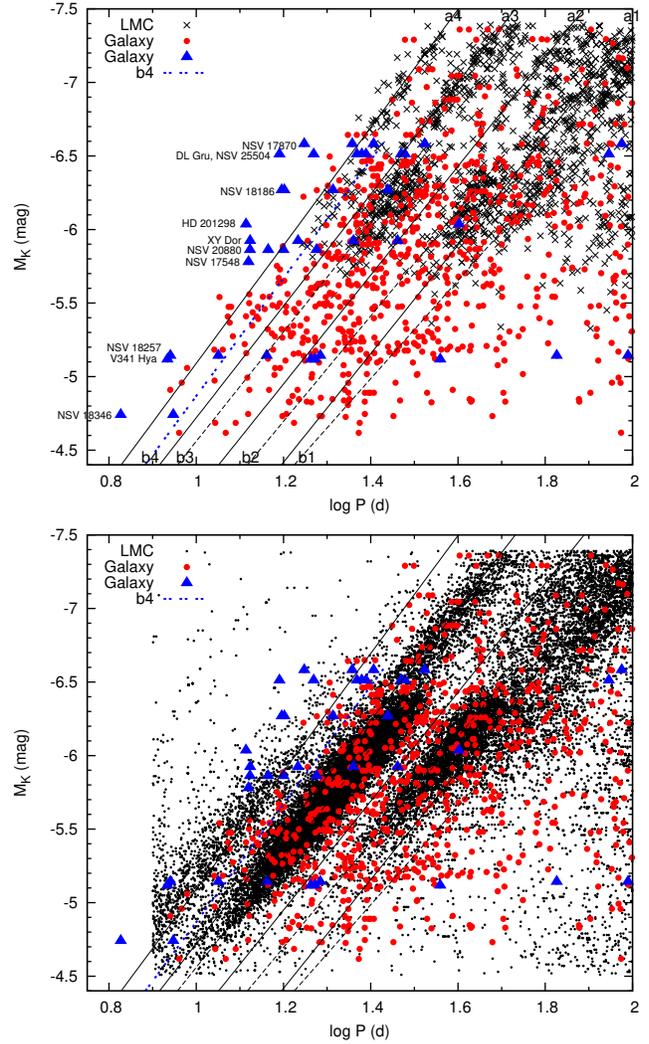}
 \caption{\vtbold{Top:} P-L sequences for local SRVs with $\sigma_{\pi}/\pi \le 0.25$ (red circle) and LMC (MACHO) periods from D06 (black cross). Sequences from S07 are shown, together with $b4$ identified by this study.  \vtbold{Bottom: Same as top panel, but using LMC(OGLE) periods.} Stars with periods shorter than $a4$ are indicated by blue triangles, including their counterparts on other longer-period sequences. These stars trace out the edge of the acoustic cut-off frequency.}
 \label{fig013}
\end{figure}

The Galactic sequences extend to \vtbold{similar low luminosities and short periods ($M_K \sim -4$ and $P \sim 10$ d) as seen in the OGLE data for the LMC (bottom two panels of Figure~\ref{fig001}), with good correlation}. Figure \ref{fig013} shows a close-up of the lower-left region of the P-L diagram \vtbold{using MACHO and OGLE periods in the top and bottom panels, respectively.}
%A low-luminosity extension of an apparent sequence extends along the left edge to $M_K \sim -4.8$ and $P \sim 6.7$ d, tracing out the limit on high-order radial pulsations imposed by the acoustic cut-off frequency \citep{b_woo06}.
\vtbold{The sequence extending along the left edge to $M_K \sim -4.8$ and $P \sim 6.7$ d, traces} out the limit on high-order radial pulsations imposed by the acoustic cut-off frequency \citep{b_woo06}.
Sequences are shown using the nomenclature in S07. Blue triangles show all periods associated with 11 stars having at least one period significantly shorter than $a4$, which corresponds to Sequence A$'$. These stars have large uncertainties in $M_K$, limiting the possibility of unique ridge identification.
%\vtbold{but appear to be associated with $a4$, based on scatter of the OGLE data in the bottom panel.}

% Fig 3.26, 082

%%%%%%%%%%%%%%%%%%%%%%%%%%%%%%%%%%%%%%%%%%%%%%%%%%%%%%%%%%%%%%%%
\section{LMC sequences: MACHO vs OGLE}
\label{s_lmc}
%%%%%%%%%%%%%%%%%%%%%%%%%%%%%%%%%%%%%%%%%%%%%%%%%%%%%%%%%%%%%%%%

To test for any metallicity dependence in the zero-point of the P-L relations, we wish to compare our data to the LMC, whose metallicity is approximately half solar: $Z_{\sun}=0.016$ and $Z_{\rm LMC}=0.008$ \citep{b_vas}. As a first step, we compared the location of the LMC P-L ($K_s$) relations derived by S07 using OGLE-II and OGLE-III photometry, with the MACHO sequences from D06 (Figure~\ref{fig013}, top panel).

To improve precision, S07 derived linear P-L relations using least-squares fits to density maps of the more numerous OGLE periods which, they noted, produced slopes that were steeper than would have been obtained using linear fits. They identified four sequences above the TRGB, labelled $a1$--$a4$, that all extend below the TRGB, and three sequences, $b1$--$b3$, that were only populated below the TRGB. The slight period shift that occurs at the TRGB \citep{b_kis03,b_ita04} arises from the temperature difference between first- and second-ascent red giants, and implies that stars below the TRGB consist mostly of RGB stars. Notably, S07 described Sequence $a4$ as containing AGB stars only, based on an earlier analysis of OGLE-II data by \citet{b_sos04}, who found no evidence for a period shift at the TRGB.

Sequence $a4$ corresponds to Sequence A$'$ in D06 and, above the TRGB, we see good agreement in Figure~\ref{fig013} with the periods derived from MACHO data. However, the extension of $a4$ below the TRGB does not match MACHO periods, which are shifted by $\delta \log P \sim -0.05$ from $a4$.
The MACHO data clearly show the expected shift at the TRGB, consistent in magnitude and direction with the shift between the longer-period sequences, implying that Sequence $``b4"$ does exist. Moreover, in Section \ref{s_pl_zp}, we confirm the presence of Sequence $b4$ in our sample of Galactic SRVs.

% Fig 3.18, 80 (originally) => 080\v6\x.eps
\begin{figure}
 \includegraphics[scale=1.0, angle=0]{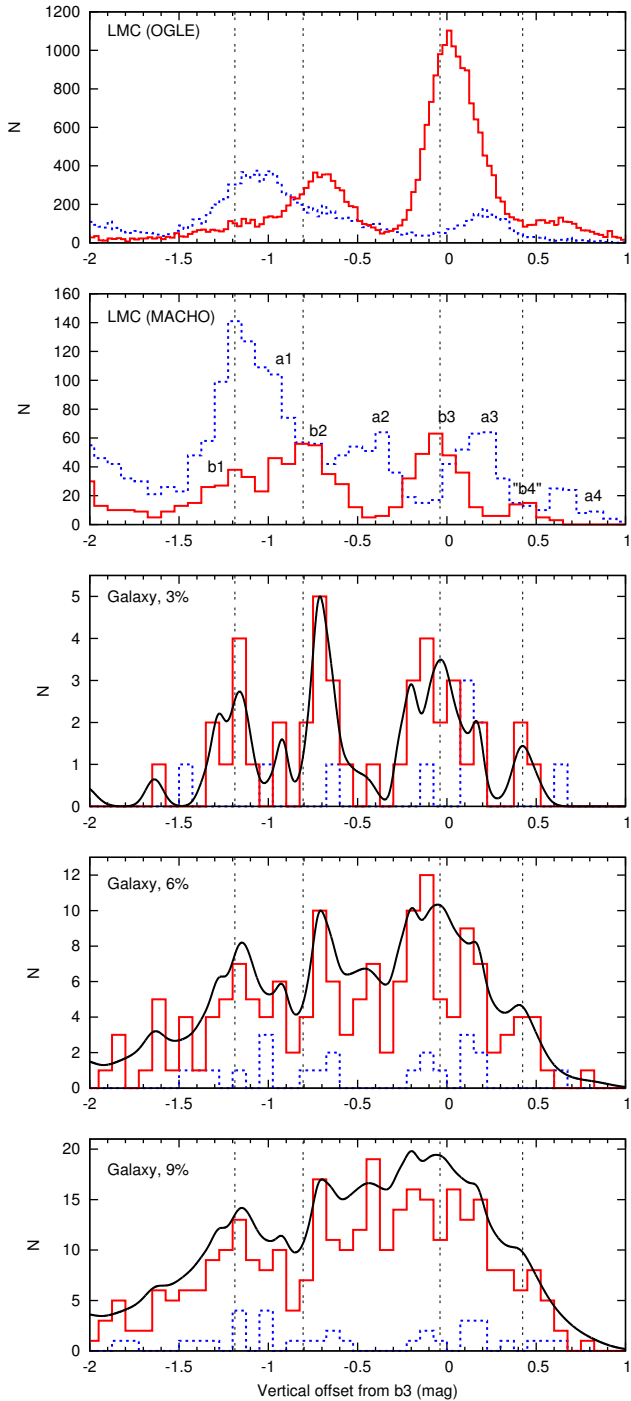}
 \caption{Vertical offsets of stars relative to $b3$, above (blue dotted) and below (red solid) the TRGB. The top panel shows LMC OGLE stars from S07, binned in 0.025 mag increments. All other panel are binned in 0.075 mag increments. The second panel shows LMC MACHO stars from D06. The remaining panels show the distributions for Galactic SRVs with $\sigma_{\pi}/\pi <$ 0.03, 0.06, and 0.09, respectively\vtbold{, assuming $\mu_{\rm LMC}=18.50$}. The vertical dotted lines indicate the best-fit positions of Sequences $b1$--$b4$ to LMC (MACHO) data, and the solid black line shows a continuous distribution function, as explained in the text.}
 \label{fig005}
\end{figure}

An alternate representation of the LMC sequences is shown in Figure \ref{fig005}, where we compare the distribution of periods as a function of vertical offset from a reference sequence, chosen here to be $b3$ \vtbold{(defined as $m_{K_s} = -3.72 \log P + 17.66$ by S07)}. The top panel shows OGLE periods from S07, selected with a variable S/N threshold, as explained in Section \ref{s_ruler}, and binned in units of 0.025 mag. The second panel contains MACHO periods from D06, binned by 0.075 mag. The bottom 3 panels show distributions for our Galactic sample \vtbold{assuming $\mu_{\rm LMC}=18.50$,} and are described in the next section. Peaks in the distributions indicate the positions of the sequences. Dotted vertical lines show the best-fit positions of Sequences $b1$--$b4$ for MACHO data, which were determined by binning the data multiple times, using a variety of bins sizes and starting offsets. There is a small disparity between peak positions in the two datasets. Although Sequences $b2$ and $b3$ are clearly identified in both datasets, OGLE peaks are shifted by about $+0.1$ mag and $+0.04$ mag, respectively, relative to MACHO. Both $b1$ and $b4$ are ill-defined in the OGLE data, with relatively broad distributions, but are clearer in MACHO. The centre of the $b1$ distribution is approximately aligned with MACHO, but $b4$ is offset by $\sim +0.2$ mag.

\begin{figure}
 \includegraphics[scale=1.0, angle=0]{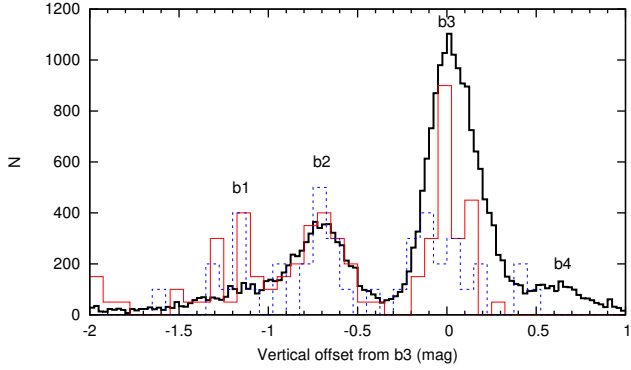}
 \caption{The OGLE and MACHO sequences below the TRGB (thick black and thin red lines, respectively) are closely aligned when the MACHO sample is restricted to the overlapping region of the LMC bar. The peaks are nearly coincident with the Galactic distribution ($\sigma_{\pi}/\pi < 0.03$, blue dashed line), which is shown for comparison, assuming $\mu_{\rm LMC}=18.50$. The MACHO and Galactic distributions have been multiplied by factors of 50 and 100, respectively, to show the relative sizes of the peaks.}
 \label{fig006}
\end{figure}

\begin{figure}
 \includegraphics[scale=1.0, angle=0]{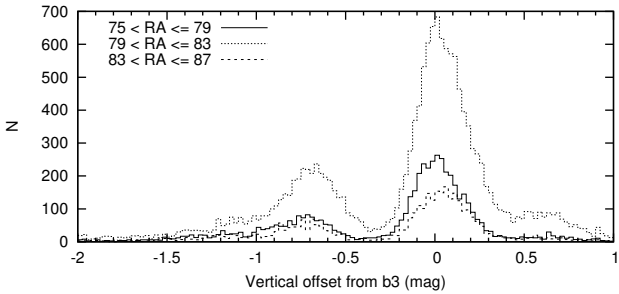}
 \caption{Subsets of the OGLE data divided into 3 regions spanning $4^{\circ}$ of right ascension along the LMC bar, binned in units of 0.025 mag. The peak offsets indicate a systematic error of $\sim 0.03$--$0.05$ mag occurs in the relative distance modulus, due to the tilt of the bar.}
 \label{fig007}
\end{figure}

\begin{figure}
 \includegraphics[scale=1.0, angle=0]{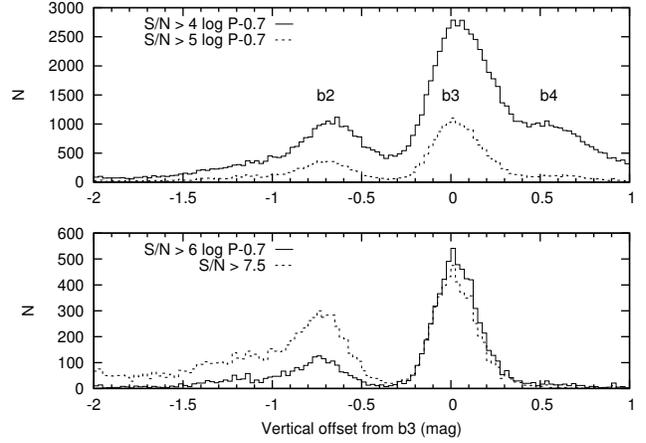}
 \caption{The distribution of stars in OGLE sequences for fixed and variable S/N thresholds. Top: A change in relative frequency \emph{and} offset of $b4$ relative to $b2$ occurs when the threshold is low enough to include low-amplitude stars on $b4$. Bottom: A comparison of variable and fixed S/N cuts, showing the change in the number of stars in $b2$ relative to $b3$, due to the different responses to the rising noise floor at longer periods.}
 \label{fig008}
\end{figure}

We investigated reasons for the discrepancy which, left unsolved, would limit possible conclusions about the effect of metallicity. Several sources of systematic error were found to contribute.
\begin{enumerate}
\item Sample selection: The OGLE survey was restricted to the LMC bar, while MACHO observed a larger area, including regions outside the bar. To test for a selection effect, we obtained the original MACHO periods\footnote{We selected stars classified as LPVs from http://macho.anu.edu.au/star} corresponding to the subset of stars overlapping the OGLE sample. The $0.1$ mag difference for Sequence $b2$ is no longer evident, with Sequences $b1$--$b3$ closely aligned with their OGLE counterparts (Figure~\ref{fig006}). The sparsely populated sequence $b4$ was not detected. Gaussian fits to the peaks show differences in their centres of $0.01$ mag for $b1$ and $b2$, and $0.03$ mag for $b3$. Although the MACHO sample contains fewer stars (its distribution has been multiplied by a factor of 50 in Figure~\ref{fig006}), the relative peak sizes approximate that of the larger OGLE sample.
\item Systematics within the LMC bar: A small systematic error is introduced by sampling along the bar, which is inclined $\sim30^{\circ}$ to our line of sight, with the eastern end being closest \citep{b_sub03,b_lah05}. To quantify this, we divided the OGLE sample into 3 groups, each spanning $4^{\circ}$ of RA, and calculated distributions for each group (Figure~\ref{fig007}). Gaussian fits %to the most populous peak ($b3$) showed a maximum systematic error of $\sim$0.03 mag in the relative distance modulus, while $b2$ showed a slightly larger difference of 0.05 mag.
    to the peaks showed variations in relative distance modulus of $0.03$--$0.05$ mag.
    A sample consisting of a mixture of stars evenly distributed along the bar will tend to smooth out this effect.
\item S/N thresholds: We selected samples from OGLE using several variable S/N thresholds, which improve the resolution of the sequences (as explained in Section \ref{s_ruler}), as well as a constant value of $7.5$, used in Paper I. Sequence $b3$ showed a $+0.03$ mag systematic shift as more stars were included, while $b2$, which contains fewer stars, showed a shift of $+0.08$ mag (Figure~\ref{fig008}). The location of $b4$ was very sensitive to the threshold. Using a variable threshold of $4 \log P - 0.7$, stars with periods on Sequence $b4$ had an effective S/N threshold of $\sim 3$, equivalent to the fixed S/N limit used by D06 to analyse MACHO data. This resulted in a shift of $\sim -0.15$ mag compared to higher thresholds (which were unable to resolve the low-amplitude stars on this sequence), and explains the large difference between the MACHO and OGLE positions for this sequence.
%101\centres\get_og.log
\item The frequency of stars occupying sequences may not be universal and may introduce a systematic error, particularly when using distribution-dependent analysis techniques, such as cross-correlation. In addition, choices made during processing, such as the S/N threshold, affects the distribution of stars within sequences. Figure~\ref{fig008}, top panel, shows a large change in frequency of $b4$ relative to $b2$ for different variable thresholds, while the bottom panel shows the effect of using fixed and variable cuts on Sequences $b2$ and $b3$. We show that these moderate changes in distribution produce relatively small systematic errors in Sect. \ref{s_ruler}.
\item Number of periods used: Restricting the number of period used for each star to a maximum of 1, 2, 3, 5 or 15 periods had a negligible effect on the positions of $b2$ and $b3$, which varied by less than $0.01$ mag.
\end{enumerate}
We conclude that with appropriate sample selection and processing, the original MACHO and OGLE periods produce consistent results, and adopt the larger OGLE sample as being representative of the LMC sequences for the remainder of this analysis.

% Fig 101\x.eps

% Fig 102\x.eps

% Fig 101\x6.eps (was x3)

%%%%%%%%%%%%%%%%%%%%%%%%%%%%%%%%%%%%%%%%%%%%%%%%%%%%
\section{Effect of metallicity on P-L zero point}
\label{s_pl_zp}
%%%%%%%%%%%%%%%%%%%%%%%%%%%%%%%%%%%%%%%%%%%%%%%%%%%%

The slopes of LMC and Galactic P-L relations have been shown to be very similar \citep{b_fou,b_gla09}. Thus, a systematic difference in peak positions between the corresponding sequences would be interpreted as a difference in the zero-points of their P-L relations. To investigate the possibility, we show the Galactic M-giant distributions for stars with $\sigma_{\pi} / \pi \le 0.03$, $0.06$ and $0.09$, respectively (bottom three panels of Figure \ref{fig005}). To take into account the spread in the photometric and parallax uncertainties, we calculated a continuous distribution function for stars below the TRGB (solid black line) as a sum of Gaussians
\begin{equation}\label{eq:1}
\Phi(o) = \sum_{i=1}^N {\frac{1}{\sigma_i \sqrt{2\pi}} \exp\left[
    -\frac{(o_i - o)^2}{2\sigma_i^2} \right] },
\end{equation}
where $o_i$ and $\sigma_i$ are their offsets relative to $b3$ and their uncertainties, respectively, and $N$ is the total number of periods in the sample \citep[]{b_sak96}. The agreement between the histograms and functions is good, with peaks closely aligned. Sequences $b1$, $b2$, $b3$, and $b4$ (vertical dotted lines) are clearly identified at 3\% and 6\% relative parallax uncertainty, but less well-defined at 9\%, due to blurring of the sequences.
In light of our findings in the previous section, we fitted Gaussians to the peaks of the Galactic sequences ($b1$, $b2$ and $b3$) and compared them to OGLE sequences defined by periods with $S/N > 7.5$, the same criterion used in Paper I. The difference in peak centres (OGLE $-$ Galactic) ranged between $-0.04$ and $+0.07$ mag, and $-0.08$ and $+0.09$ mag for $\sigma_{\pi} / \pi \le 0.03$ and $0.06$, respectively. The mean differences for the two samples were $+0.03 \pm 0.05$ and $+0.01 \pm 0.07$ mag, respectively, indicating identical peak centres to within the uncertainties.

We used cross-correlation to assess the similarity of the OGLE and Galactic distributions below the TRGB, where the majority of periods were detected. A discrete cross-correlation function,
\begin{equation}\label{eq:2}
R_{xy}(m) = \sum_{n=0}^N  x[n] \; y[n+m]
\end{equation}
was applied, where $x$ and $y$ are the binned LMC and Galactic distributions, respectively, $N$ is the number of bins, and $m$ is the trial offset. The best match is indicated by the value of $m$ that maximises $R$. Since Figure~\ref{fig005} shows the distributions are closely aligned, only values of $m$ that implied an offset less than 0.25 mag were searched. $R$ was calculated for a range of bin sizes and starting offsets to reduce sensitivity to these parameters, yielding a mean offset and the rms scatter, which we adopted as our uncertainty.
% - cd F:\ccd\phot\mgiant\use\090404a
% - doit 0.10 2 (range of plxs)
% - vizie\crosscor.bat
We found offsets implying that the Galactic distributions were fainter than their OGLE counterparts by $0.04$ and $0.07$ mag for samples with $\sigma_{\pi} / \pi \le 0.03$ and $0.06$, respectively, with uncertainties of $\sim 0.03$ mag. The expected $K$-band zero-point shift for an increase in metallicity by a factor of 2 is $\sim 0.13$ mag \citep{b_woo90}, which is supported by observational evidence from a comparison of SMC and LMC sequences \citep{b_ita04}. Although the Galactic shift is in the expected direction, its magnitude is significantly smaller than expected and is inconsistent with our previous result based upon peak centres. We caution that cross-correlation assumes an identical morphology of the signals being compared, which is not the case for the distribution of variables occupying the sequences in the two galaxies (Figure~\ref{fig006}).
\vtbold{On the other hand, these comparisons assume $\mu_{\rm LMC}=18.50$ and, a small change in the adopted value would systematically alter the alignment of the distributions, potentially affecting the conclusion.}
We use a novel new method in the following section to investigate further.

%%%%%%%%%%%%%%%%%%%%%%%%%%%%%%%%%%%%%%%%%%%%%%%%%%%%
\section{Estimating absolute magnitudes from periods}
\label{s_ruler}
%%%%%%%%%%%%%%%%%%%%%%%%%%%%%%%%%%%%%%%%%%%%%%%%%%%%

In order to use pulsating M giants as distance indicators, we now describe a new method that utilises the periods alone to determine the absolute magnitude of a given star, using the P-L sequences of the LMC as the reference.

Paper I confirmed the multi-periodic nature of local M giants and highlighted the similarity of their pulsation properties to their counterparts in the LMC. Following the evidence presented in Sect. \ref{s_pl_zp}, we hypothesise that the LMC P-L sequences are representative of sequences in the SMC and local solar neighbourhood, with identical slopes and zero-points. Using the distribution of LMC stars in the ($m_{K_s}$, $\log P$) plane at specific periods, we derive probability density functions that describe the likelihood of a star having a particular magnitude. Using all measured periods for a given star, we estimate its magnitude by multiplying these probability density functions and locating the peak in the resulting probability function.

We constructed a fiducial LMC P-L diagram using OGLE-II and OGLE-III periods from S07. $K_s$-band magnitudes were extracted from the Infrared Survey Facility catalogue (IRSF; \citealt{b_kat07}), which provides deeper and more accurate photometry than 2MASS, due in part to its superior resolution. To reduce the possibility of false matches, only stars that had a single match within a 1" radius were used. We transformed the IRSF magnitudes to the 2MASS system using the relation $K_{\rm s,IRSF} = K_{\rm s,2MASS} + 0.010 (J-K_s)_{\rm 2MASS} + 0.014$ \citep{b_kat07} in order to compare the magnitudes, and found good agreement. However, we noted a rapid divergence at faint magnitudes for stars with $0.7 \le \log P < 0.9$, which we excluded from the analysis.

\begin{figure}
 \includegraphics[scale=1.0, angle=0]{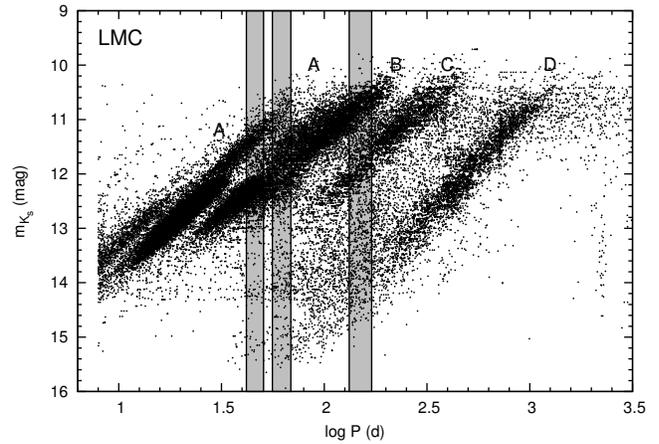}
 \caption{LMC P-L diagram using OGLE periods and IRSF magnitudes. Periods were selected using an adaptive S/N cut-off which helped to resolve the shorter-period sequences at low S/N. Grey vertical bars correspond to the parts of the P-L used to create the probability density functions shown in Figure \ref{fig032}.}
 \label{fig031}
\end{figure}

% F:\MSC\OSARG\PRO\OSARG.AWK
We combined OGLE data for OSARGs, Mira, SRV, and ellipsoidal variables into one dataset, while limiting the periods to the range $0.9 \le \log P \le 3.5$. Due to the rising noise-floor caused by $1/f$ noise, we departed from the usual practise of using a fixed S/N cut-off. Instead, we used a sloping noise floor when calculating the S/N of each peak, rejecting periods with $S/N < 5 \log P - 0.7$. This allowed only high S/N periods (at large P) to be selected, while still allowing short-periods with lower S/N to be used (limited by an additional constraint of $S/N \ge 3$). This scheme resulted in good resolution of all known sequences (Figure \ref{fig031}).

% Fig 098/v2/x1.eps - OGLE P-L

To estimate a star's magnitude, we first created a histogram of the stellar density within a narrow, vertical strip of the P-L diagram ($\pm 2.5$\% wide in $\log P$), centred about each of its periods. To provide adequate resolution, counts were made for bins of 0.01 mag, which were then smoothed over $\pm 5$ bins using a boxcar average. Each resulting distribution, which is basically the luminosity function of stars having that pulsation period, was converted to a probability density function by normalising the area under the curve to unity. In this way, the function describes the probability of a star with that period having a particular magnitude. By combining multiple periods by multiplying the corresponding probability density functions, an estimate of the magnitude of the star is given by the bin with the highest relative confidence.

\begin{figure}
 \includegraphics[scale=1.0, angle=0]{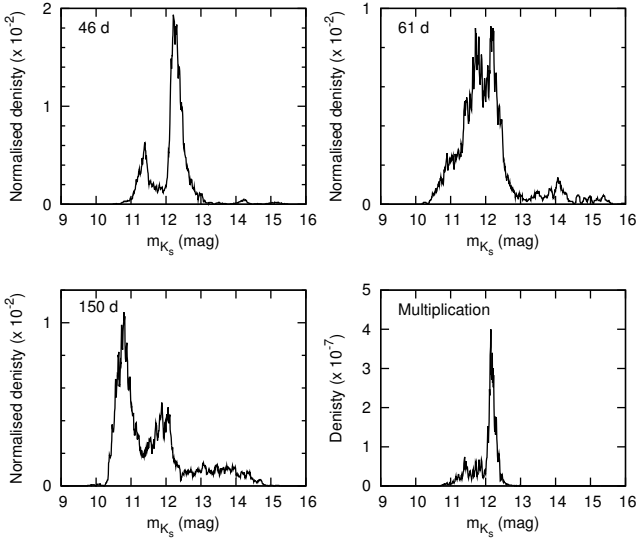}
 \caption{Probability density functions corresponding to the vertical strips in Figure \ref{fig031}. The combined probability (bottom right) indicates the most likely magnitude as $m_{K_s} = 12.15$, in close agreement with the measured value of $12.18 \pm 0.02$ mag.}
 \label{fig032}
\end{figure}

As an example, we consider a star in the LMC with 3 measured periods: 46 d, 61 d, and 150 d. The corresponding vertical strips centred at $\log P = 1.66$, $1.79$ and $2.18$ are shown in Figure \ref{fig031}. The normalised probability functions are shown in Figure \ref{fig032}, together with the combined probability, which indicates an estimated magnitude of $m_{K_s} = 12.15$. This is in close agreement with the measured value of $m_{K_s} = 12.18 \pm 0.02$.

% Fig 098/v2/x2.eps  - period cuts and density

\begin{figure}
 \includegraphics[scale=1.0, angle=0]{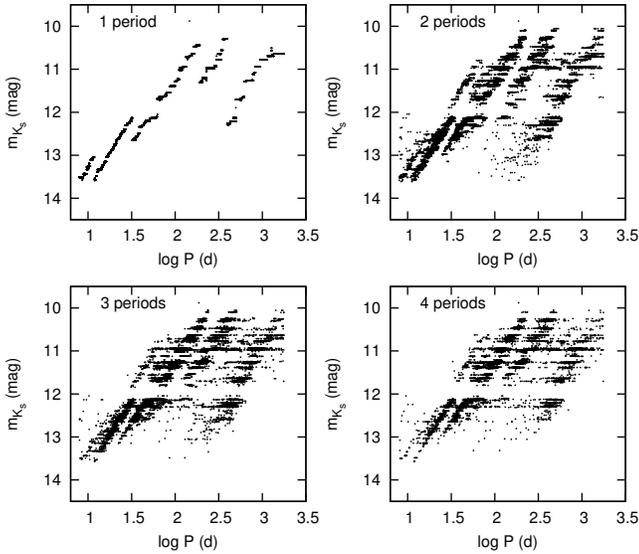}
 \caption{LMC P-L diagrams using estimated magnitudes calculated using 1--4 of the most significant periods for each star. The resolution of the sequences improves as more periods are used to estimate magnitudes.}
 \label{fig033}
\end{figure}

As a further test, we used the procedure to recreate the LMC P-L diagram using estimated magnitudes. The OGLE periods for each star used to produce Figure \ref{fig031} were sorted by descending pulsation amplitude. Magnitudes were estimated using the above technique using the $n$ most significant periods, where $n=1$, 2, 3 and 4 (Figure \ref{fig033}). A single period is sufficient to reproduce the main sequences, although the estimated mag at a given period always corresponds to the peak of the probability function and, thus, stars may be allocated to incorrect sequences. Adding further constraints (more periods) produces more reliable results, with the estimates no longer falling on the most populated sequence. Using 2 periods, Sequence $A$ above the TRGB is resolved, and Sequences $A'$ and $B$ are more accurate. As the number of periods increases, a smoothing in the density of the P-L occurs, with gaps being filled in. We note that periods corresponding to faint magnitudes are moved upward in the diagram, due to the algorithm's propensity to gravitate toward areas of greater density in the P-L plane and, thus, it does not reproduce the known magnitudes in this case. This effect also explains the exaggerated gap at the TRGB.

% Fig 098/v2/x3.eps - P_Ls

\begin{figure}
 \includegraphics[scale=1.0, angle=0]{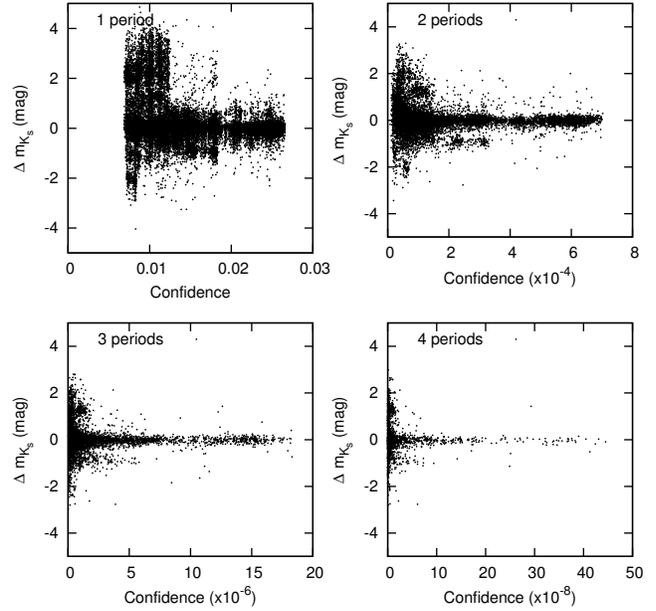}
 \caption{LMC confidence measures as a function of magnitude difference (estimated $-$ measured). The scatter decreases as more periods are used to estimate magnitudes.}
 \label{fig034}
\end{figure}

Figure \ref{fig034} shows the difference between the measured and estimated magnitudes, plotted against the confidence measure of each estimate, calculated as the height of the peak in the combined probability distribution. The confidence is arbitrary, in the sense that it cannot be compared with other values produced with a different number of periods. It is only useful as a relative measure to compare estimates produced using the same number of periods. Figure \ref{fig034} confirms that magnitudes estimated from a single period have a significant chance of allocating stars to incorrect sequences, but results improve quickly as more periods are used to constrain the estimate. Estimates using 2, 3 and 4 periods produce progressively fewer outliers and a narrow distribution centred about zero, indicating that the algorithm is working well.

% Fig 098/v2/x4.eps - confidence LMC

\subsection{The SMC distance modulus}

We now apply this method to derive the distance modulus of the SMC relative to the LMC. As before, we obtained periods from OGLE-II and III (S07), and matched the SMC stars to the IRSF catalogue to obtain $K_s$-band magnitudes. Magnitudes were estimated using the 2 or 3 strongest periods (largest amplitude), since sample sizes were significantly reduced when requiring more periods satisfying our adaptive S/N cut-off.

\begin{figure}
 \includegraphics[scale=1.0, angle=0]{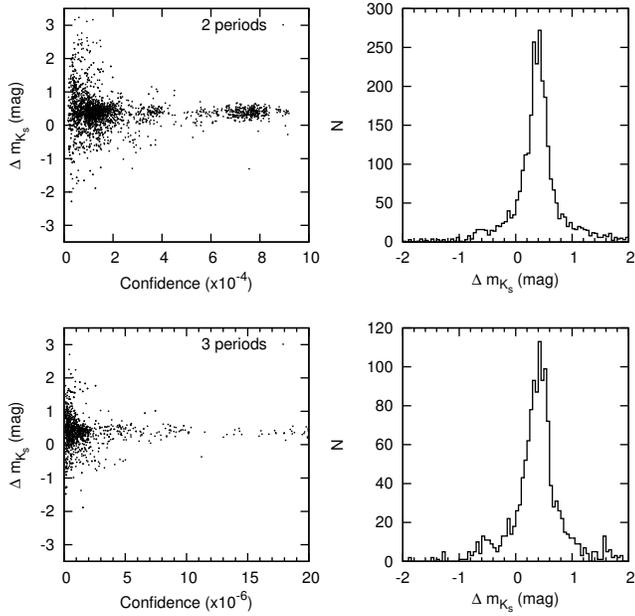}
 \caption{Magnitude differences plotted against confidence (left panels) for SMC stars using 2 and 3 OGLE periods, respectively, and the corresponding distributions (right panels). The offset of $\sim 0.40$ mag represents the SMC distance modulus relative to the LMC.}
 \label{fig035}
\end{figure}

We used an iterative approach to account for the different luminosity ranges in the two galaxies. An initial pass using all periods yielded an approximate relative distance modulus of $\Delta \mu \approx 0.4$ mag. Using this value, a second pass omitted periods that did not overlap the luminosity range of the LMC (specifically, those that were fainter than the 95th percentile of the distribution at a given period), resulting in an improved estimate and a smoother distribution, with fewer outliers.
The left-hand panels of Figure \ref{fig035} show the magnitude difference plotted against the confidence. Clearly, the difference is non-zero, allowing us to measure the distance modulus of the SMC relative to the LMC.

% Fig 098/v2/x8.eps - SMC - was x5

% 098\v2\smc2.bat
The right-hand panels show the magnitude differences binned in units of 0.05 mag. It should be emphasised that the distribution is not a true Gaussian, since the wings contain some stars that were allocated to incorrect sequences. However, the core of the peak is composed of stars with high confidence levels that were likely to be assigned to the correct sequence. Gaussian fits to the distributions yield a relative SMC distance modulus of $\Delta \mu = 0.39 \pm 0.01$ mag and $0.40 \pm 0.01$ mag for 2 and 3 periods respectively, which are consistent with the values derived from the subset of stars with high confidence levels only. The quoted uncertainties are the formal uncertainties from the Gaussian fit.

To assess the sensitivity of this method to systematic errors introduced by different S/N thresholds and, indirectly, the relative number of periods in each sequence, the samples were reprocessed using periods selected with the same fixed and variable S/N thresholds shown in Figure~\ref{fig008}. The results were very similar, with a mean relative distance modulus of $\Delta \mu = 0.40 \pm 0.01$ mag and $0.41 \pm 0.02$ mag for 2 and 3 periods, respectively.
Using an extensive list of distances published since 2002 to calculate a mean distance modulus for the LMC and SMC, \citet{b_tam08} found a difference of $0.40 \pm 0.02$ mag, in excellent agreement with our result.

\subsection{Calibration of the LMC distance modulus using Hipparcos parallaxes}

We now apply the algorithm to our sample of nearby Galactic M giants to give the magnitude of each star relative to the LMC fiducial. Paper I listed all periods derived from our Fourier analysis, which included some closely spaced frequencies that were probably caused by amplitude and phase modulations of a single oscillation mode. To remove them, all periods for a given star were sorted by descending amplitude. Each period was then compared to the remaining periods (of lower amplitude) to remove those with a period ratio $< 1.1$, which were assumed to be phase/amplitude modulations of the primary period. Periods longer than 150 d were also excluded, to avoid spurious periods caused by $1/f$ noise.

% Fig 098/v2/x7.eps - Gal (2MASS) 2P
\begin{figure}
 \includegraphics[scale=1.0, angle=0]{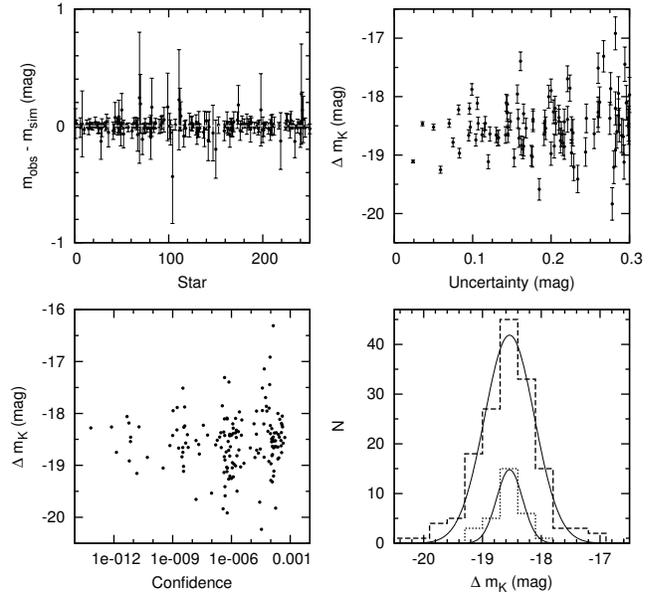}
 \caption{Galactic M giant results using 2--5 periods. Top left: difference between magnitude estimates derived from the observed periods and a Monte-Carlo simulation. Magnitude differences ($M_{K}$ - estimated) as a function of uncertainty (top right) and confidence (bottom left). Bottom right: distribution of estimated magnitudes for the full sample (dashed) and the subset with uncertainties $< 0.15$ mag (dotted).}
 \label{fig037}
\end{figure}

To improve the statistics, we used all stars in the sample having 2--5 periods, regardless of their relative parallax and photometric uncertainties. We propagated uncertainties in the observed periods, assumed to be 3\%, to estimated magnitudes using a Monte-Carlo simulation with 10000 iterations for each star. In most cases, the estimated magnitudes derived from the observed periods were close to the simulation values (Figure \ref{fig037}, top left). Stars with large uncertainties were found to have either broad peaks or multiple peaks of similar height in their probability function, which yielded disparate estimates from only small changes to the input periods. We show the combined uncertainties from period, parallax and photometric uncertainties in Figure \ref{fig037}, top right.

%As noted earlier, faint stars do not reproduce the known magnitudes correctly and, since our Galactic sample extends to lower luminosities than the OGLE data for the LMC, 
We used two passes to determine the LMC distance modulus.
Due to the use of a variable number of periods per star, confidence values are not easily comparable, although a logarithmic plot shows four distinct groups, indicating the number periods used (Figure \ref{fig037}, bottom left). The distribution of magnitude offsets with respect to the LMC is shown in the bottom right panel, binned by 0.3 mag (dashed line). The distribution for the subset of stars with uncertainties smaller than 0.15 mag is shown with a dotted line. Gaussian fits to the two distributions (solid lines), yield an identical distance modulus of $\mu_{\rm LMC} = 18.54$ mag, with a formal uncertainty of $\sim 0.02$ mag. Reprocessing our Galactic sample with OGLE periods selected with the S/N thresholds shown in Figure~\ref{fig008} yielded a relatively small range of distance modulus spanning $0.08$ mag, with a mean of $\mu_{\rm LMC} = 18.54 \pm 0.03$ mag. Tests using the subset of stars with the smallest uncertainties produced nearly identical results. Our distance modulus agrees with the mean value of $18.53 \pm 0.01$ mag derived from 16 recent, independent, distance determinations \citep{b_tam08}, which is identical to within the uncertainties. Thus, our data show that the P-L relation zero-points have a negligible metallicity dependence.

As a final test of this method, we reprocessed our sample with OGLE periods selected from stars within the three R.A. ranges shown in Figure~\ref{fig007}, in an attempt to detect a systematic shift caused by the tilt of the LMC bar. This yielded values of $\mu_{\rm LMC} = 18.59$, $18.54$ and $18.52 \pm 0.01$ mag for ranges centred on $R.A. = 77^{\circ}$, $81^{\circ}$ and $85^{\circ}$, respectively. These distances correctly show that the eastern end of the bar is closest, while the range in relative distance modulus agrees with the recently determined value of $0.1 \pm 0.03$ mag \citep{b_lah05}.

%Our method has several advantages over the Mira P-L relationship for determining distances. SRVs are more numerous and, having faster variability, they require less monitoring to measure their pulsation periods. Their multi-periodic nature places constraints on their absolute magnitude, allowing a single star to be used to estimate distance. Miras are more luminous and have higher amplitudes, placing less demands on photometric precision. However, when viewed in isolation, they may be confused with stars on the LSP sequence, leading to incorrect inferences on distance. Unlike short-period SRVs, Miras are more affected by circumstellar extinction due to dust, which introduces an additional source of uncertainty. Our method makes use of both Mira and SRV sequences and, having a negligible metallicity dependence, is well-suited to measuring distances in the solar neighbourhood, Galaxy and beyond.

\vtbold{
Our method has some advantages over the Mira P-L relationship for determining distances. SRVs are more numerous and, their multi-periodic nature places constraints on their absolute magnitude, allowing a single star to be used to estimate distance. Miras are more luminous and have higher amplitudes, placing less demands on photometric precision. %Although SRVs have faster variability, they require monitoring over similar time scales as the longer-period Miras, due to their smaller amplitudes and complex multi-periodic behaviour.
Both methods require monitoring over similar time scales to measure either one long period or, multiple shorter periods.
% as the longer-period Miras, due to their smaller amplitudes and complex multi-periodic behaviour.
%
%However, they may be confused with SRVs, which may also occupy Seq. C (Woo99), although the smaller amplitudes of SRVs ($K_s < 1.5$ mag) are sufficient to identify them (Mat09). %Similarly confusion , However, when viewed in isolation, they may be confused with stars on the LSP sequence, leading to incorrect inferences on distance, although careful photometric monitoring should find indications of the shorter-period.
However, unlike short-period SRVs, Miras are more affected by circumstellar extinction due to dust, which can be significant at periods longer than $\sim 125$ d \citep{b_ala01}, and introduces an additional source of uncertainty.
Our method makes use of both Mira and SRV sequences and, having a negligible metallicity dependence, is well-suited to measuring distances in the solar neighbourhood, Galaxy and beyond.
% LSP has short and long period w/PR of ~ 5-13.  wood 99
% Fig 16 of Matsunaga suggest A>1.5 mag to eliminate SRVs
}

%%%%%%%%%%%%%%%%%%%%%%%%%%%%%%%%%%%%%%%%%%
\section{Pulsation Amplitudes}
\label{s_Amp}
%%%%%%%%%%%%%%%%%%%%%%%%%%%%%%%%%%%%%%%%%%

The usefulness of M giants as distance indicators relies upon RGB pulsation properties being universal. Paper I showed that both period ratios and the incidence of multi-periodicity in nearby M giants are consistent with stars in the Galactic Bulge and Magellanic Clouds. We now examine the pulsation amplitudes of nearby M giants, to better understand the limitations, if any, they impose on our understanding of the P-L sequences. It is also interesting to study pulsation amplitudes in their own right, for example, to determine the relative roles of Mira-like and solar-like excitation processes (see Sect. \ref{s_SLO}).

\begin{figure}
 \includegraphics[scale=0.7, angle=0]{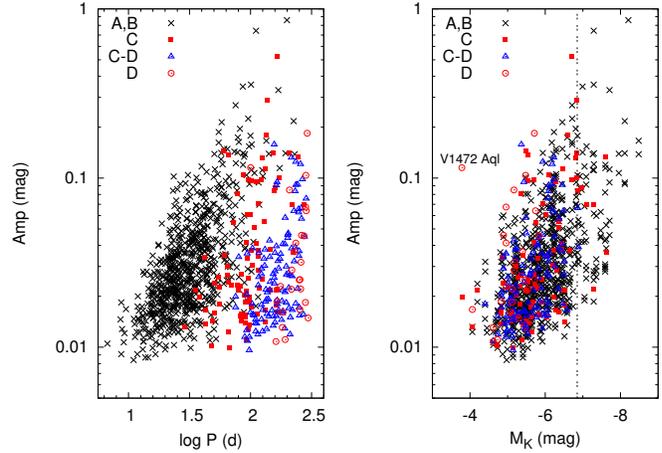}
 \caption{Period-amplitude (left) and luminosity-amplitude (right) relations for our survey fields. CCD amplitudes are shown in the instrumental system, which has  a broad response that peaks between $V$ and $R$ bands. Squares denote periods on Sequence C of \citet{b_woo99}, which contains stars pulsating in the fundamental mode, while crosses are used for those pulsating in overtones. Open circles denote stars on seq D, containing stars with Long Secondary Periods, and triangles denote stars between Sequences C \& D.}
 \label{fig014}
\end{figure}

Our survey detected peak-to-peak Fourier amplitudes ranging from 0.86 mag ($\pi^1$ Gru) %K257
down to 8 mmag ($\phi^2$ Hya), % K229
with 90\% of periods having amplitudes $A < 0.1$ mag, and 75\% having $A < 0.05$ mag. % 1069/1178, 892/1178
We show period-amplitude and luminosity-amplitude diagrams for stars with $\sigma_{\pi} / \pi \le 0.25$ in Figure~\ref{fig014}. Different symbols have been used for periods corresponding to fundamental-mode pulsation (labelled C), higher overtones (labelled A,B), long secondary periods which have unknown origin; see, for example, \citealt{b_woo04,b_nic09} (D), and those falling between Sequences C and D (labelled C-D). Periods were assigned to particular sequence when they were within $\pm 0.5$~mag of the relations defined by S07 (Table 1). Data from \citet{b_gla07} were omitted since amplitude data were not available for all published periods.

In the period-amplitude diagram (left panel), the majority of detected periods correspond to short-period overtone pulsation. A clear trend of increasing pulsation amplitude with period is evident, including a fairly sharp upper envelope. \citet{b_ala01} studied SRVs in the low-extinction region of Baade's window of the Galactic Bulge, finding that amplitude increased with period, reaching a typical value of about 0.3 mag at 100\,d, in agreement with our results. Similarly, \citet{b_wra} identified 15000 red variables in the Galactic bar, with a typical $I$-band amplitude of about 0.2 mag at 100 d. An increase in pulsation amplitude with luminosity, and therefore in period, is expected for solar-like oscillations \citep{b_kje95}. The observations for overtone pulsators (A and B) lend support to the idea that pulsations in these stars are excited stochastically by convection.

Sequence C shows a weaker correlation between period and amplitude, in agreement with the results of D06 for the LMC. Sequence D contains very few periods, but a positive correlation with amplitude is also visible. Notably, the periods falling between Sequences C and D, which we have attributed to low-frequency noise (see Sect. \ref{s_sample}), are positively correlated with amplitude and have a slope similar to the other sequences. This result suggests a physical origin and can be explained by stellar noise from convection, which is expected to increase towards lower frequencies. OGLE detected a significant number of periods between Sequences C and D (see Figure \ref{fig031}), consistent with our result.

The luminosity-amplitude diagram (right panel of Figure~\ref{fig014}) shows that pulsation amplitudes are positively correlated with absolute luminosity for all sequences. A sudden decrease in density occurs around $M_K \sim -6.85$ mag (dashed line), which marks the TRGB \citep{b_tab09b}. With the exception of a few outliers, the majority of pulsation-related amplitudes below the TRGB are smaller than 0.2 mag, in agreement with the period-luminosity-amplitude data-cube slices of the LMC and SMC, which show very few stars with $A > 0.14$ mag \citep{b_kis03,b_kis04}.

% Fig 3.27, 071/x.png

%%%%%%%%%%%%%%%%%%%%%%%%%%%%%%%
\subsection{V1472 Aquilae}
%%%%%%%%%%%%%%%%%%%%%%%%%%%%%%%

\begin{figure}
 \includegraphics[scale=1.0, angle=0]{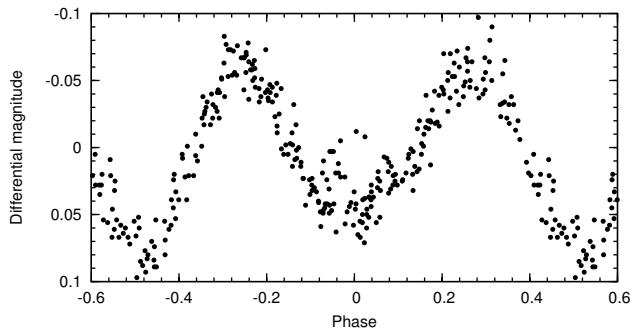}
 \caption{Phased light curve of V1472 Aql folded with $P$=200.05 d. The continuous brightness variations and primary and secondary minima are typical of an ellipsoidal variable.}
 \label{fig021}
\end{figure}

Stars in the luminosity-amplitude plot obey a well-defined amplitude limit at a given luminosity, with the exception of V1472 Aql, whose amplitude is unexpectedly large for its absolute magnitude ($M_K=-$3.78, $A$=0.12 mag, $P$=100.03 d), suggesting that this period is not pulsation related. Our period and amplitude determinations are in close agreement with the Hipparcos values of $P$=100.37 d and $A_{\rm Hp}$=0.16 mag, both of which place the star on the low-luminosity extension of Sequence D, which contains ellipsoidal variables (D06). \citet{b_luc82} used 41 radial velocity measurements to determine a spectroscopic period of 198 d, and concluded that it could be a contact binary that may show eclipses. Indeed, using the spectroscopic period, \citet{b_sam97} folded the Hipparcos photometry to plot a low-quality light curve revealing continuous brightness variations, and possible indications of primary and secondary minima, consistent with an eclipsing or ellipsoidal variable star. %Processing our photometry with iterative sinewave fitting \citep{b_fra} to remove the primary period from its power spectrum, we find a low-amplitude secondary period of $\sim$66 d  (Figure~\ref{fig020}). The period-ratio of exactly 1.5 implies that V1472 Aql exhibits two alternating amplitudes separated by a period of 100 d, which are clearly
Our data show that V1472 Aql exhibits the primary and secondary minima of an ellipsoidal system (Figure~\ref{fig021}).

\begin{comment}  ############
% Fig 3.6, 067/235/235ac.eps
\begin{figure}
 \includegraphics[scale=1.0, angle=0]{fig020}
 \caption{Power spectra of V1472 Aql, showing primary and secondary periods of 100.025 d and 66.683 d (dotted and solid lines, respectively). The amplitude difference and period ratio of 1.500 imply a light curve that alternates between two slightly different amplitudes every 100 d, consistent with primary and secondary maxima of an eclipsing binary system. The inset shows the window function.}
 \label{fig020}
\end{figure}
\end{comment}  ############

% Fig 067/235/235ph.eps

%%%%%%%%%%%%%%%%%%%%%%%%%%%%%%%%%%%%%%%%%%%%%%%%%%
\section{Bridging the gap to G and K giants}
\label{s_SLO}
%%%%%%%%%%%%%%%%%%%%%%%%%%%%%%%%%%%%%%%%%%%%%%%%%%

The most luminous AGB stars undergo stable, mono-periodic Mira-like pulsations that are driven by opacity changes. On the other hand, solar-like oscillations, the stochastically driven excitation and damping of linearly stable modes by convection \citep{b_dzi} may explain the less regular light variations of SRVs \citep{b_chr}. Examples of solar-like oscillations have been identified amongst Galactic SRVs \citep{b_bed03}, including $L_2$ Pup, which exhibits characteristics of both Mira-like pulsation and solar-like oscillations \citep{b_bed05}. Low-amplitude, short-period photometric variability has also been observed in G and K giants, with amplitudes of 5--15 mmag and periods of 2--4 d \citep{b_edm,b_hen}, leading to the conclusion that pulsation may extend to spectral classes earlier than M. However the precise nature of the driving mechanism is still unknown.

\begin{figure}
 \includegraphics[scale=1.0, angle=0]{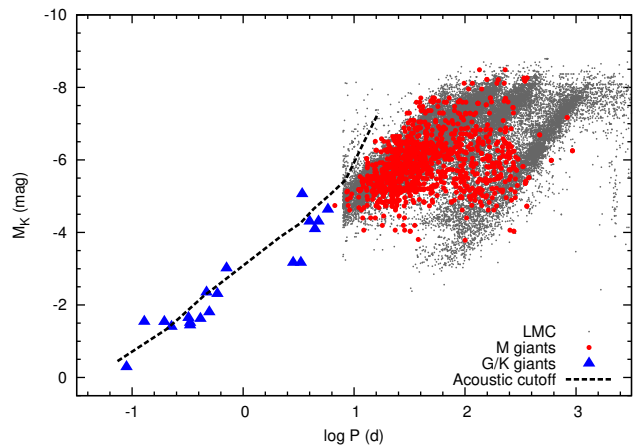}
 \caption{P-L diagram extended to include nearby Galactic G and K giants exhibiting solar-like oscillations. A dashed line marks the approximate location of the acoustic cutoff frequency, scaled from the solar value of 5.5 mHz, as explained in the text.}  % SRVs to 25%
 \label{fig016}
\end{figure}

Recently, solar-like oscillations have been observed in a number of G and K giants with well determined parallaxes, allowing them to be added to our P-L diagram. Table \ref{T3} summarises the results from 7 recent papers, tabulating the Hipparcos catalogue number, star name, spectral type, revised Hipparcos parallax and uncertainty \citep{b_van07}, apparent $K$ magnitude, absolute $K$ magnitude, magnitude source (2=2MASS, D=DIRBE, G=Gezari, m=mean of all three) as discussed in Paper I, period corresponding to the frequency of maximum power, and a reference to the original paper. Additionally, data for 11 K giants observed with the WIRE satellite were taken from Table 1 of \citet{b_ste}.
Periods in hundreds more G and K giants have been measured by the space missions CoRoT \citep{b_hek09} and Kepler \citep{b_bed10}. However, with the exception of HR~7349, a lack of parallaxes prevents them from being included in Figure~\ref{fig016}.

% Table 3.7
\begin{table}
  \caption{G/K giants with relatively precise parallaxes that exhibit solar-like oscillations. References:
  A=\citet{b_fra02}, B=\citet{b_ret03}, C=\citet{b_tar07}, D=\citet{b_tar08}, E=\citet{b_derid06}, F=\citet{b_bar04}, G=\citet{b_car10}.}
  \label{T3}
  \setlength{\tabcolsep}{1.5mm}
{\scriptsize
  \begin{tabular}{@{}rlrrrrlll}
  \hline
  HIP   & Name           & Sp	& $\pi$        & $m_K$   & $M_K$ & Src & Period     & Ref \\
        &                &      & (mas)        & (mag)   & (mag) &     & (d)        &     \\
  \hline
  56343 & $\xi$ Hya      & G7   & 25.16 (0.16) & +1.46   & $-1.54$ & G & 0.13       & A   \\
  69673	& $\alpha$ Boo   & K1.5 & 88.83 (0.54) & $-$2.91 & $-3.17$ & 2 & 2.82, 3.30 & B,C \\
  72607 & $\beta$ UMi    & K4   & 24.91 (0.12) & $-$1.29 & $-4.31$ & 2 & 3.96, 4.74 & D   \\
  79882 & $\epsilon$ Oph & G9.5 & 30.64 (0.20) & +1.03   & $-1.54$ & m & 0.19       & F,E \\
  89962 & $\eta$ Ser     & K0   & 53.93 (0.18) & +1.05   & $-0.29$ & 2 & 0.09       & F   \\
  95222 & HR 7349        & G8   &  9.64 (0.34) & +3.46   & $-1.62$ & 2 & 0.41       & G   \\
  \hline
\end{tabular}
}
\end{table}

The combined P-L diagram (Figure~\ref{fig016}) shows the LMC sequences from \vtbold{S07} (grey circles), our local SRVs with $\sigma_{\pi}/\pi \le 0.25$ (red circles) and the G and K giants (blue triangles). 
%Our Galactic data extend to
\vtbold{The M giants have} sufficiently short periods and low luminosities to bridge the gap to the most luminous, longest period G and K giants.

The left-most edge of the M-giant sequences traces the acoustic cutoff frequency, whose approximate location has been plotted by scaling the Sun's frequency of $\nu_{\rm ac} \sim 5.5$ mHz using the relation $\nu_{\rm ac} \propto g T_{\rm eff}^{-1/2} \propto M R^{-2} T_{\rm eff}^{-1/2}$ \citep{b_bro91}, using stellar parameters from \citet{b_BOB} and $V-K$ colours from \citet{b_bes}.

The emerging picture is a clear progression of properties as we ascend the giant branch. Stochastically excited, low-amplitude, micro-variability in G and K giants progresses to a mixture of Mira-like pulsation and solar-like oscillations in SRVs, and ends with stable, large-amplitude, Mira-like pulsation. Our data nicely bridge the gap between G/K and M giant pulsation, demonstrating that a smooth continuity exists.

% Fig 3.29, 083

%%%%%%%%%%%%%%%%%%%%%%%%%%%%%%%%%%%%%%
\section{Conclusion}
\label{s_conclusion}
%%%%%%%%%%%%%%%%%%%%%%%%%%%%%%%%%%%%%%

We have conducted a 5.5-yr survey of \vtbold{243} southern, pulsating M giants in the local solar neighbourhood. Combining our data with additional literature periods, we have analysed a sample of 286 stars with relatively precise Hipparcos parallaxes to unambiguously detect 4 P-L sequences below the TRGB. Together with Paper I, we have shown that the pulsation-related properties of nearby M giants, including period ratios, amplitudes, multi-periodicity, and the temperature related shift at the TRGB are consistent with SRVs found in the Magellanic Clouds and Bulge, indicating that pulsation in a truly universal and consistent phenomenon. The metallicity dependence of the P-L zero points are negligible over the metallicity range spanned by the SMC, LMC and Galaxy, indicating that SRV P-L relationships are suitable as high-quality distance indicators. A new method for estimating absolute magnitudes from pulsation periods is described, which yields an LMC distance modulus of $\mu_{\rm LMC} = 18.54 \pm 0.03$ mag, calibrated using Hipparcos parallaxes. We demonstrate that V1472 Aql is an ellipsoidal variable, whose period and absolute magnitude place it on the low-luminosity extension of Sequence D, in agreement with earlier conclusions about the nature and corrected location of Sequence D. We show that 17 nearby G and K giants that exhibit solar-like oscillations have periods similarly constrained by the acoustic cutoff frequency. Our data bridge the gap, both in luminosity and period, between G and K giant solar-like oscillations, and M giant pulsation, demonstrating a clear continuity as we ascend the giant branch.

%%%%%%%%%%%%%%%%%%%%%%%%%%%%%%%%
\section*{Acknowledgments}
%%%%%%%%%%%%%%%%%%%%%%%%%%%%%%%%

We thank Igor Soszy\'nski for supplying OGLE-III data, and the referee for helpful comments and suggestions. This research has made use of the VizieR and SIMBAD databases, operated at CDS, Strasbourg, France. This project has been supported by the Australian Research Council, the Lend\"ulet Young Researchers Program of the Hungarian Academy of Sciences and the Hungarian OTKA Grants K76816 and MB0C 81013.

%%%%%%%%%%%%%%%%%%%%%%%%%%%%%%%%

\bsp

\label{lastpage}

\end{document}